\documentclass{article}
\usepackage[utf8]{inputenc}
\usepackage{graphicx}
\usepackage{geometry}
 \usepackage{mdframed}

\usepackage{amsmath}
\usepackage{amssymb}
 \usepackage{url}

\usepackage{xcolor}
\usepackage[normalem]{ulem}

\usepackage{epstopdf}
\usepackage{subfigure}

\usepackage[square,numbers]{natbib}
\begin{document}

\title{Physical grounds for causal perspectivalism.}
\author{G. J. Milburn, S. Shrapnel and P. W. Evans}

\maketitle

\begin{abstract}
    We ground the asymmetry of causal relations in the internal physical states of a special kind of open and irreversible physical system, a causal agent. A causal agent is an autonomous physical system, maintained in a steady state, far from thermal equilibrium, with special subsystems: sensors, actuators, and learning machines. Using feedback, the learning machine, driven purely by thermodynamic constraints, changes its internal states to learn probabilistic functional relations inherent in correlations between sensor and actuator records. We argue that these functional relations just are causal relations learned by the agent, and so such causal relations are simply relations between the internal physical states of a causal agent. We show that learning is driven by a thermodynamic principle: the error rate is minimised when the dissipated power is minimised. While the internal states of a causal agent are necessarily stochastic, the learned causal relations are shared by all machines with the same hardware embedded in the same environment. We argue that this dependence of causal relations on such `hardware' is a novel demonstration of causal perspectivalism.
\end{abstract}

\section{Introduction}

Is causation in the external, physical world or in our heads? Russell~\cite{Russell1913} famously denied the former while the latter seems unacceptably subjective. The interventionist account of causation~\cite{Pearl2000,Woodward2003}, especially when interpreted along perspectival lines~\cite{Ismael2015}, seems to be somewhere in between, ``an irenic third way" in the words of Price~\cite{Price2007}. We demonstrate here a plausible physical schema within which causal claims depend for their truth on the internal physical states of a special kind of machine, a causal agent. We take this to be an exemplification of a perspectival view of causation that is not anthropocentric, and is dependent on the laws of physics, especially thermodynamics. Our objective is to give empirical support to Price's ``causal viewpoint" as ``a distinctive mix of knowledge, ignorance and practical ability that a creature must apparently exemplify, if it is to be capable of employing causal concepts''~\cite{Price2007}. In Cartwright's terms, we physically ``ground the distinction  between effective strategies and ineffective ones''~\cite{Cartwright1979}. Or as Ismael writes ``Causes appear in mature science not as necessary connections written into the fabric of nature, but robust pathways that can be utilized as strategic routes to bringing about ends''~\cite{Ismael}. We seek these `robust pathways' in the physical structure of learning machines. 

We take it as given that causal concepts find no application in the fundamental physics of closed systems. This is, we think, what Russell~\cite{Russell1913} must have had in mind. A system is closed if it cannot interact with anything outside itself. The physical dynamical laws of a closed system are reversible in time. In contrast, how we define an open system depends on where we draw the boundary. We will call a system `open' if its internal interactions are much stronger than any interactions with systems external to it. Of course, such a characterisation is highly contextual, and is dependent on, say, the nature of our experimental access to the system and the time-scale of such access: in the long run, even small interactions will matter. The dynamical laws that describe open systems are irreversible and stochastic. It is our contention that casual relations can only be understood in terms of open systems.

We describe a special kind of open system -- a causal agent (CA) -- an open system maintained in a {\em non-thermal-equilibrium} steady state. A CA contains specialised subsystems: sensors, actuators, and learning machines. The external world does work on the CA through special subsystems called {\em sensors}. The CA does work on the world through special subsystems called {\em actuators}. There is an essential thermodynamic asymmetry to sensors and actuators: in both cases, the operation of the subsystem dissipates energy. The learning machine is an irreversible physical system that exploits functional relations inherent in correlations between sensor and actuator states in order to optimise the use of thermodynamic resources. These `causal relations' become embodied in the bias settings of the learning machine. As above, learning machines are necessarily dissipative and noisy~\cite{CP-LM,Seifert2017}.

\begin{figure}
    \centering
    \includegraphics[width=0.8\textwidth]{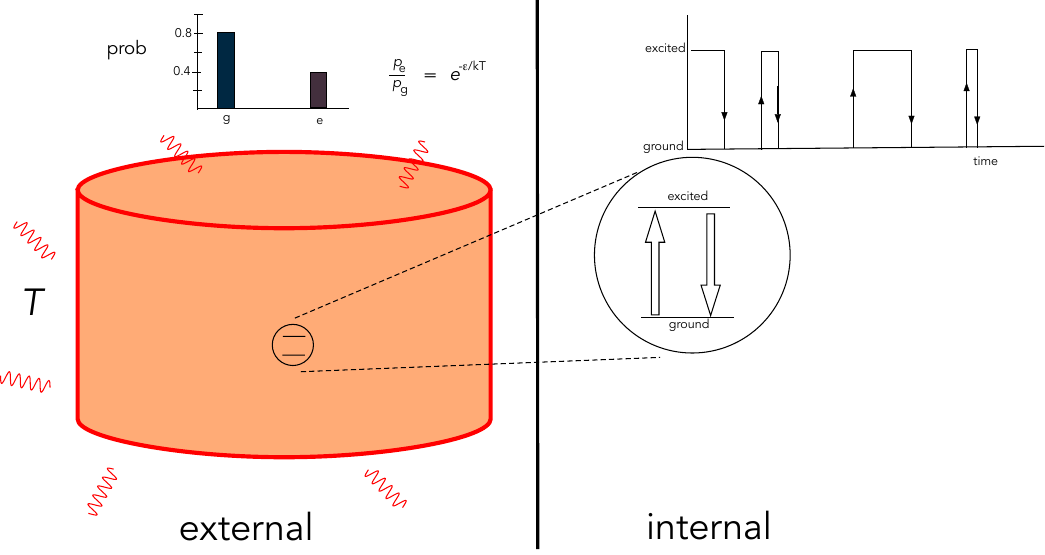}
    \caption{A depiction of the distinction between the external and internal description of an open system: a two-level atom interacting with a thermal radiation field inside a cavity at temperature $T$. On the left is the external description in terms of the stationary probability distribution to find the atom in the ground state ($g)$ or excited state ($e$). On the right is the stochastic record of the continuously observed state of a single atom as a function of time.}
    \label{in-out}
\end{figure}

We make a distinction between two ways one can describe the irreversible dynamics of an open system, which we call {\em internal} and {\em external} descriptions. The internal description of an open system is a fine-grained description that provides a complete specification of the values of the internal physical states of a single system as a function of time. The external description of an open system arises as a result of coarse-graining over the internal physical states, and thus gives only the probability distribution of internal physical states or, equivalently, average values of physical quantities of the internal description, for example the average energy in terms of temperature or average number of particles in terms of chemical potential. While the probability distributions of the external description can be stationary, the internal description of physical states is stochastic. For example, a single two-level atom interacting with a thermal radiation field at fixed temperature may be described in terms of its stationary Boltzman distribution --- the external description -- or in terms of a stochastic switching between its internal states at rates determined by the temperature of the environment, see Fig.(\ref{in-out}).

The distinction we are making maps onto the equivalent ways of describing diffusion given by Langevin (internal) and Einstein (external). We wish to make clear that the distinction between the internal and external descriptions of a CA is not the same as the distinction between subjective and objective. The internal and external descriptions both refer to objective physical states as seen from a `third person' perspective with different levels of grain, and different epistemic access, to those states.  

One of our primary goals in this work is to show how causal relations can be defined in terms of the physical states of the learning machine of the CA. In terms of the internal description, the stochastic record of sensations and actions of the CA are coupled by feedback to the irreversible process of the learning machine. The effect is to drive the learning machine stochastically to a new state in which various internal settings are fixed by implicit correlations between sensor and actuator records. The internal settings (weights and biases) of the learning machine continue to fluctuate once the leaning is complete, but nonetheless they enable prediction of the sensations that follow an action with little probability of error. On this view the causal relations {\em just are} the internal settings of the learning machine.\footnote{We might imagine that these internal settings are simply the modelling parameters of the causal model constructed by the CA.} 

In the external description, the probability distribution over the internal settings of the learning machine evolve by feedback to a new stationary distribution centred on particular domains in the state space of the learning machine. While the probability distributions over sensor and actuator records can remain completely random, the probability distribution of the settings of the learning machine are `cooled' to particular steady state distributions that correspond to the learned causal relation implicit in the correlations between sensor and actuator records. We use the term `cooled' to emphasise that the dynamics of the learning machine is dissipative. It lowers its entropy at the expense of increasing the entropy of its environment. 

Our argument proceeds as follows. In \S\ref{simple-model} we discuss a very simple classical dynamical model of a causal agent and illustrate the distinction between the internal and external descriptions of the CA.\footnote{Our construction of a causal agent has considerable overlap with similar constructs by other authors. The importance of sensors and actuators for artificial agents is a staple of textbooks on artificial intelligence~\cite{RusNor2010} that provide the raw data for learning algorithms. Briegel~\cite{Briegel} also stresses the importance of sensors and actuators for embodied agents. His novel concept of `projective simulations' plays the role of a learning machine in our model. He emphasises the role of stochasticity for creative learning agents and possible quantum enhancements. In elucidating his concept of action based semantics Floridi~\cite{Floridi2011} describes a two-machine artificial agent (AA). The `two machines' of his scheme roughly correspond to the actuator/sensor machines and the learning machine in our model. Our model of a prediction/correction learning machine is close to the agent model introduced in a biological setting by Friston~\cite{Friston} and subsequent developments~\cite{Bruineberg}. Freeman~\cite{Freeman} links causation directly to learning and motor processes in central nervous systems.} In \S\ref{sensor-actuators} we discuss the thermodynamics of sensors and actuators and the fundamental asymmetry between them. In \S\ref{learning-machine} we introduce the concept of a learning machine (not a machine learning algorithm!) as an irreversible physical system coupled by feedback to the sensors and actuators. We give a simple example of how a learning system can embody causal concepts in its physical steady states. In \S\ref{therodynamics-learning} we discuss the thermodynamic constraints on learning machines and in \S\ref{learning-causes} we discuss how learning machines based on prediction and correction feedback can learn causal relations. In \S\ref{conclusion} we make explicit how our model of learning in a CA is an exemplification of causal perspectivalism.

\section{A simple physical model}
\label{simple-model}

In this section we begin to develop our model of a causal agent (CA) using a simple example.   The objective of the example  is to introduce the three essential components of a CA --- sensors, actuators, learning machine --- in terms of a  simple physical system. We first give the external description of each component followed by the internal description. This highlights key thermodynamic processes and irreversibility. The learning machine component can be described both in terms of a physical feedback process and in terms of learning a causal functional relation.  

\subsection{The external description}

\begin{figure}[t]
    \centering
    \includegraphics[scale=0.6]{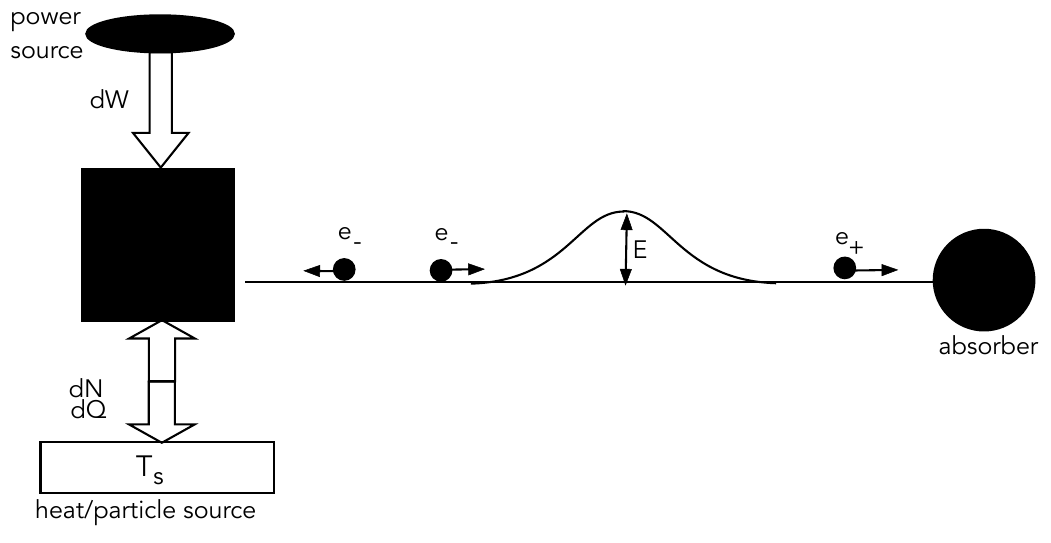} 
    \caption{The system on the left, the source, is connected to a source of work (increment $dW$) and also a thermal/particle reservoir (increments $dQ,dN$ respectively).  When the work done on the source is  zero it is in thermal equilibrium and simply heats a distant absorber. By doing work on the source we can bias it to emit predominantly low energy particles and reduce the heating rate of the absorber.}
    \label{bump}
\end{figure}

Consider the following simple classical dynamical system (Fig.(\ref{bump})). A local system, the `source', can emit particles of variable kinetic energy. It is driven by an external power supply towards a non-equilibrium steady state by dissipating power into a local environment at temperature $T_s$. The emitted particles travel towards a small potential hill. If they have enough kinetic energy, they can surmount the hill and never return to the source. If they do not have sufficient kinetic energy they will be reflected from the potential hill and return to the source. We will assume that the motion of the particles once they leave the source is entirely conservative, that is to say, particles move without friction.  As particles that surmount the barrier are lost, we will assume that the source is supplied by a particle reservoir such that the average particle number of the source is constant in time. But to  make very sure that particles with sufficient energy to surmount the barrier do not encounter a yet higher barrier we will add a particle absorber to the right of the potential hill. Overall, this entire physical arrangement is an irreversible system.

Let us begin by assuming that no work is done on the particle source. Suppose the source initially emits particles which can take one of two possible values of kinetic energy $(e_-,e_+)$, with equal probability, such that  $e_- < E$ aand $e_+ > E$.  All particles that have energy $e_+$ are absorbed at the right thereby raising the internal energy of the absorber. All particles that have energy $e_- < E$ are returned to the system and absorbed. The system has access to an environment that emits/absorbs  particles locally to keep the average particle number and average energy constant.  The entropy of the source is one bit in natural units and its average energy $(e_-+e_+)/2$. The average energy of the absorber is steadily increasing as it absorbs particles from the source. This energy is heat extracted from the source's environment. 

According to the external description of the system, the source is simply a source of heat and particles. Since the laws of thermodynamics tell us that there are no perfect absorbers, we know that eventually the absorber will start emitting something -- even if ultimately it becomes so massive that it undergoes gravitational collapse to form a black hole and emit Hawking radiation. We can, however, make the general assumption that the absorber is at a lower temperature than the source, and so heat transferred from the source to the absorber raises the overall entropy. As a result, the total system is not in thermal equilibrium. The external description can thus be completely specified in terms of the total average number of particles $\bar{N}(t)$ available at any time and the average number of particles of each species as distinguished by energy $\bar{n}_{\pm}(t)$ corresponding to energies $e_{\pm}$.

Assume that the CA operates in time steps of duration $T$. IN each time step it emits an $e_+$ particle with probability $p_+$ and an $e_-$ particle with probability $p_-$. For simplicity we will assume that these are the mutually exclusive events that can happen in each time step so that $p_++p_-=1$. This is clearly a binomial process. 

In $K$ time steps, the mean number of $e_+$ particles emitted is $Kp_+$. As all these particles are lost,  the total average energy lost is thus $\Delta\bar{E}= Kp_+$. It is clear now what needs to happen. The machine needs to reduce the probability that it emits a high energy particle, that is to say,  $p_+$ is a decreasing function of $K$.  

The external description is simple enough. The feedback modulates a power source to change the thermodynamic state of the source to lower its entropy and push it away from thermal equilibrium. This is a very simple form of feedback control, a simple kind of learning. What kind of internal mechanism would implement this? 

\subsection{The internal description}
\begin{figure}[t]
    \centering
    \includegraphics[scale=0.9]{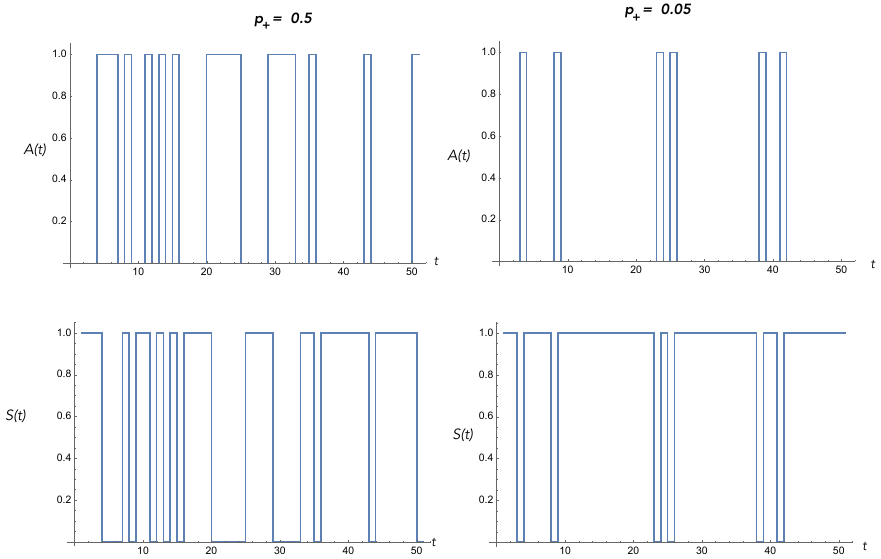}
    \caption{Examples of the stochastic record of particles emitted by the actuator and received by the sensor as a function of time for two different values of $p_+$. The time step is $T=1$.}
    \label{S-A-records}
\end{figure}

The internal description is given in terms of a history of actions and sensations.  We can give a simple model of the internal mechanism in the source in terms of sensor and actuator records. In each time step, the actuator record tracks the energy of each emitted particle:  ${\bf 0 }$ if an $e_-$ particle was emitted and ${\bf 1}$ if an $e_+$ particle was emitted. In each time step, the sensor record tracks whether a particle was received back: ${\bf 0}$ if no particle is received back and ${\bf 1}$ if a particle is received back. The logical relation of the sensations as a function of actions is binary NOT. One way that we can imagine the operation of the machine is that it is required to simulate the NOT function with small probability of error. However there is an equivalent physically motivated way to achieve the same operational behaviour.  In Fig. (\ref{S-A-records}) we plot sample trajectories for the actuator and sensor records in terms of the code defined above. 

\begin{figure}[t]
    \centering
    \includegraphics[scale=0.5]{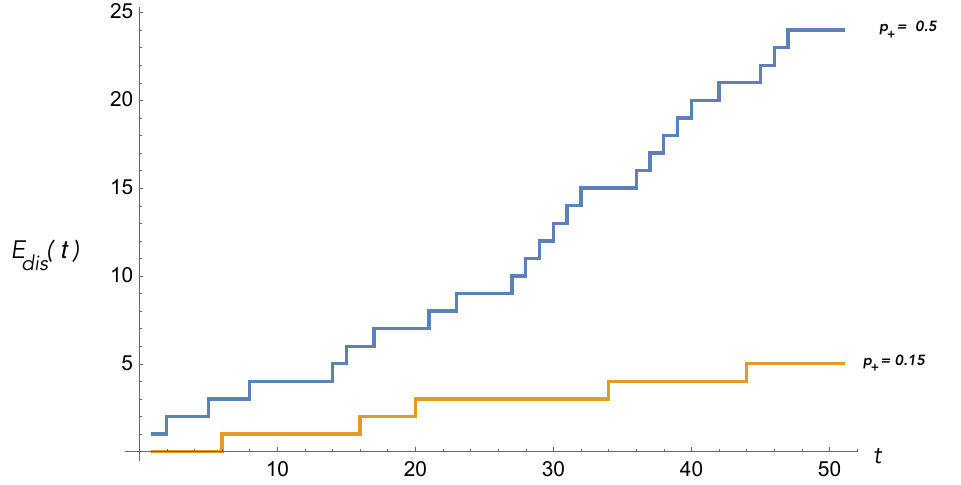}
    \caption{Two sample trajectories of the stochastic energy dissipated by the CA as a function of time for two different values of $p_+$. The time step is $T=1$.}
    \label{energy-dis}
\end{figure}

The next step is to consider how the machine is dissipating energy. Using the code we defined in the previous paragraph we can see that the total emitted energy, in a time interval $t$, is simply $e_+$ times the total number of  ${\bf 1}$'s recorded up to time interval  $t $. The energy dissipated by the machine  as a function of time is 
\begin{equation}
E_{dis}(t) =\int_0^t\ dt\ A(t')~.
\end{equation}  
This is a stochastic variable. In Fig.(\ref{energy-dis}) we plot $E_{dis}(t)$ for two sample trajectories. 

A simple internal feedback process can now be defined that will minimise the energy dissipated by the machine in each time step (the dissipated power). We let the probability to emit a high energy particle depend on time $p_+(t)$,
\begin{equation}
   p_+(t)=p_+(0)e^{-\lambda\int_0^t dt' A(t')}~,
\end{equation}
where $A(t)$ is the stochastic record of the actuator and $\lambda$ is a positive constant called the learning rate. 
This is equivalent to 
\begin{equation}
\label{stochastic-prob}
    \frac{dp_+(t)}{dt} =-\lambda A(t)~.
\end{equation}
As we have defined a discrete-time process, we can write this
\begin{equation}
    p_+(n+1)= (1-lA(n))p_+(n)~,
\end{equation}
where $l$ plays the role of the learning rate. 
This is a stochastic difference equation as the record $A(n)$ is stochastic. All learning processes in the internal description are necessarily stochastic. Clearly this is a highly non-Markov process: the entire history of the actuator records are used in the feedback.

In what sense is this kind of non-Markov feedback described as learning? The answer takes us to the connection between learning and thermodynamics, or between information and physics. The objective of the machine is to minimise energy cost operating in this environment. At each time step the sensor record as a function of the action record can be described as a binary NOT.  But the machine knows nothing about Boolean functions. All it has access to are its sensor and actuator records and an imperative  to minimise the energy dissipated in its interactions with the world. If it can find a way to do this it will have learned correlations implicit in its actuator and sensor records.  We return to this question in \S\ref{therodynamics-learning}.

In reality, both actuators and sensors have a small error probability so that the record does not exactly match what actually happened in the actuator and sensor devices. Due to such errors it is impossible to reach a state in which the sensor record is composed entirely of ${\bf 1}$s. That is an un-physical state of zero entropy. All that is required is that the probability of finding a zero in the sensor record is very small.  

This example is very simple, and so the causal law it discovers is also simple: if it emits a particle with low energy, it is most likely to be returned, otherwise not. This is the `law of physics' from the perspective of this CA. Emitting a particle is an intervention. The sensor responds accordingly and the agent learns the causal relation as a result of its intervention. The internal records of the CA are completely contingent as the inside description shows. As the actuator/sensor records are random binary strings every CA will, almost certainly, have different records. The physical internal states of each CA is unique, but according to the external description the internal records are unknown (by definition), and every CA of this type behaves as an equivalent thermodynamic machine minimising average thermodynamic cost and thereby lowering its average entropy. 

\section{Thermodynamic constraints on sensors and actuators}
\label{sensor-actuators}

In this section we will formulate general principles that capture key features of the simple example introduced in the previous section. There is a thermodynamic asymmetry between sensors and actuators. Actuators do work on the world, whereas the world does work on sensors. Work done on/by a system is constrained by the change in the free energy (Helmholtz or Gibbs). The change in the free energy constrains the work done on/by the sensors and actuators of a CA. The average work done by a CA must be less than the decrease in free energy. Physical changes to the sensors {\em increase} the free energy of the CA while actuators  {\em decrease} the free energy of the CA. These thermodynamic asymmetries must be built into the physical construction of sensors and actuators and are  a defining feature of a causal agent. 

Average work, and corresponding changes in free energy, are part of the external description of a CA. To provide the internal description of the process, we make use of the Jarzynski equality~\cite{Jarzynski}. On this account, work is a random variable conditioned on contingent physical processes inside the agent. The Jarzynski equality relates the statistics of this stochastic process to changes in free energy.  

\begin{figure}
    \centering
    \includegraphics[scale=0.7]{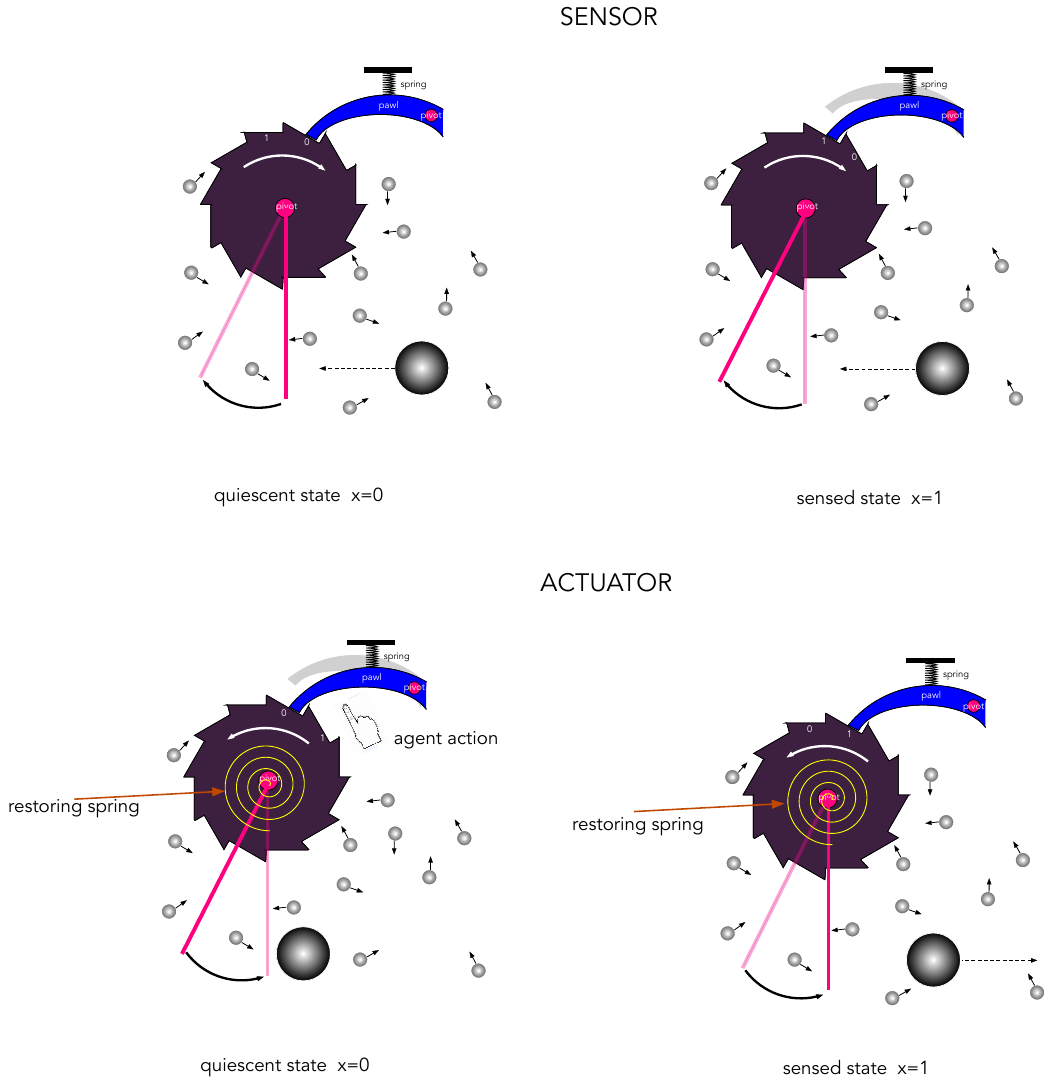}
    \caption{Example of a mechanical sensor and actuator based on a ratchet and pawl. The state of each device is represented by the binary numbers written on the ratchet. The collision of the `signal' particle is inelastic and leads to an irreversible transfer of energy to the device through a combination of elastic compression of the pawl spring and friction. The signal particle does work on the sensor. In the case of the sensor, gravity applies a continuous force to maintain the quiescent state of the pendulum arm. The continuous collision of the background particles slowly restores the quiescent state. In the case of the actuator a coiled spring in the ratchet (constantly rewound by the agent) applies a constant torque opposing gravity to maintain the quiescent locked state of the ratchet. The agent `acts' by pushing the pawl off the ratchet, the pendulum falls to eject a `probe' particle, increasing its kinetic energy and thus extracting work from gravity. The continuous collision of the background particles slowly restores the quiescent state due to the constant force applied by the restoring coiled spring inside the ratchet mechanism. The background collisions with the noise particles occasionally switches the device corresponding to an error.}
    \label{fig:mechanical-SA}
\end{figure}

A  mechanical example of the thermodynamic asymmetry of actuators and agents is given in (Fig.(\ref{fig:mechanical-SA})). The objective of the sensor is to detect a collision of a large `signal' particle, while the objective of the actuator is to eject a large `probe' particle. The dissipation and noise is represented by a large number of much smaller particles colliding with the mechanism.

Let us assume for simplicity that both actuator and sensors have only two physically distinguishable states labeled by a binary variable $x=\{0,1\}$ and that the energies of each state $E_x$ are such that $E_0<E_1$. We assume that, in the absence of actions and sensations, each system is highly likely to be found in an initial `ready' state. In the case of a sensor the ready state is the lower energy state and in the case of an actuator the ready state is the high energy state. According to the coarse-grained external description, the system is described by a probability distribution $p_x(t)$. In the internal description the system is described by the value of a binary stochastic variable $x(t)$.

 We will use a two state birth-death master equation model to give the external description of sensors and actuators. In the absence of interactions between the agent and the external world, the occupation probability for each state  is the stationary solution to the two-state Markov  process of the form
\begin{equation}
\label{two-state-me}
    \frac{dp_0}{dt}  = -\gamma_1p_0+\gamma_0 p_1=-\frac{dp_1}{dt}~,
\end{equation}
where $\gamma_1$ corresponds to a transition $0\rightarrow 1$ and $\gamma_0$ corresponds to the transition $1\rightarrow 0$. 
The corresponding stationary distributions are then given by 
\begin{equation}
\label{ss}
    \frac{p_1(\infty)}{p_0(\infty)}= \frac{\gamma_1}{\gamma_0}~.
\end{equation}
The conditions that distinguish the quiescent state  of sensors and actuators  (the `ready' state for sensations and action) are 
\begin{eqnarray}
    \gamma_1 & < & \gamma_0\ \ \ \ \ \ \ \ \ \ \mbox{(sensor)}\\
    \gamma_1 & > & \gamma_0\ \ \ \ \ \ \ \ \ \ \mbox{(actuator)}~.
\end{eqnarray}
In the case of a sensor, prior to a sensation it is more likely to be found in the lower energy state $x=0$ than the higher energy state $x=1$. In the case of an actuator, prior to an action it is more likely to be found in the higher energy state $x=1$ than the low energy state $x=0$.

It is important to stress however that neither the sensor nor the actuator are in thermal equilibrium with their environment. They are in non-equilibrium steady states due to external driving of a dissipative system, the causal agent as a whole. In classical physics the rates $\gamma_{0,1}$ go to zero as the temperature goes to zero.\footnote{In the quantum case they may not go to zero at zero temperature due to dissipative quantum tunnelling (as in optical bistability, for example \cite{Carmichael}).}

The stationary distributions give the external describe the identical systems but, as we have seen, an individual system is certainly not stationary on the internal description; it is switching between the two states at rates determined by $\gamma_{0},\gamma_{1}$. The two descriptions are connected by time-averaging. In a long time average, the ratio of the times spent in each state is given by the ratio of the transition probabilities
\begin{equation}
    \frac{\tau_1}{\tau_0}= \frac{\gamma_1}{\gamma_0}~,
\end{equation}
in the limit that the total time $\tau_1+\tau_0 \rightarrow \infty$.  Thus prior to actions and sensations, the sensor spends more time in the lower energy state $x=0$ while the actuator spends more time in the higher energy state $x=1$.

On the internal account, we are interested in describing the energy of a single system. In the current example, $x(t)$ is a stochastic process (a random telegraph signal). We define two Poisson processes $dN(t)_{x}$ ($x=\{0,1\}$) that can take the values $0,1$ in a small time interval $t$ to $t+dt$. In most small time intervals the state of the system does not change, but every now and then the system can jump from one state to the other. If a jump does happen, one or the other of $dN(t)_{x}=1$ in the infinitesimal time interval $t\rightarrow t+dt$.  The probability to take the value $1$ in time interval  $dt$ is then simply
\begin{equation}
\label{SA-bd-me}
    P(dN_x(t)=1;t+dt,t) ={\cal E}(dN_x) =\gamma_x dt~,
\end{equation}
where ${\cal E}$ defines the average. These equations imply that the continuous record of the state label $x(t)$ satisfies the stochastic differential equation, 
\begin{equation}
\label{sde-rts}
    dx(t)=(1-x(t))dN_+(t)+x(t)dN_-(t)~.
\end{equation}
The internal states of other components in the agent are responding to these fluctuating signals at all times.  The agent is said to be in a ready state or \emph{quiescent state} if the time average of the signal corresponds to the stationary states in Eq.(\ref{ss}).  

\begin{figure}[t]
    \centering
    \includegraphics[scale=0.3]{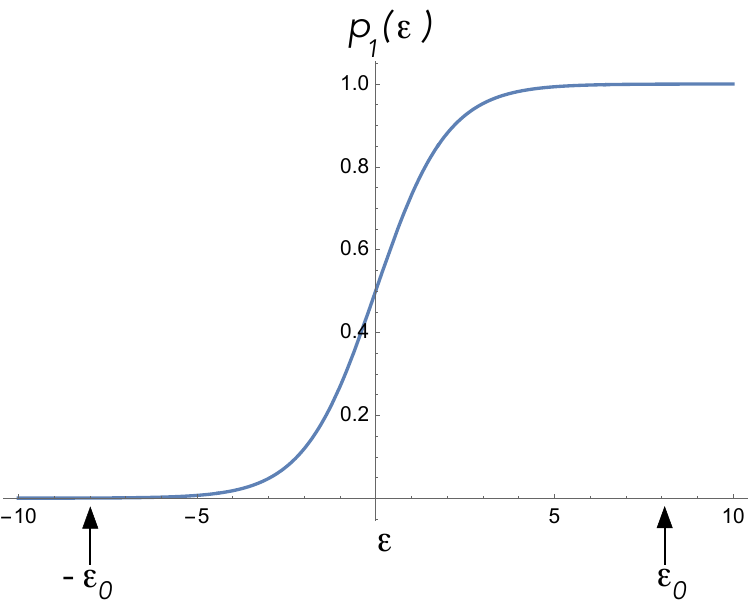}
    \caption{The probability for the sensor and actuator to be find in the high energy state as a function of a bias parameter $\epsilon$. The bias values for the quiescent states of each are indicated by $\epsilon_0< 0$ for a sensor and $\epsilon_0 > 0$ for an actuator.  When internal and external control functions act, the bias value is changed from $\epsilon_0$ to $\epsilon_f=-\epsilon_0$.   }
    \label{act-sens-bias}
\end{figure}

We now describe how these devices respond to internal (actuator) and external (sensor) inputs. In both cases bias forces act on sensors and actuators in such a way as to make the transition rates time-dependent. This is how work is done on/by the system and during which time heat is dissipated as they are pushed away from their quiescent steady-states. We will refer to these inputs as the control functions. To show this, first we define
\begin{eqnarray}
   \gamma_0 & = &\gamma e^{-\epsilon/2}\\
   \gamma_1 & = &\gamma e^{\epsilon/2}~.
\end{eqnarray}
where we refer to $\epsilon$ as the bias. 
If $\epsilon$ is a constant, the steady state average energy is
\begin{equation}
\bar{E}=E(1+e^{-\epsilon})^{-1}~,
\end{equation}
This function is plotted in Fig. (\ref{act-sens-bias}). To set the devices to their quiescent state, a particular bias value $\epsilon_0$ is chosen. In the case of a sensor $\epsilon_0 <0$ and $\bar{E}$ is small,  while in the case of an actuator $\epsilon_0 >0$ and $\bar{E}$ approaches $E$.

Physical forces inside (actions) or outside (sensations) the CA change the value of $\epsilon$ from the quiescent state bias, $\epsilon_0$, at time $t_i$ to a final value $\epsilon_f$ at time $t_f$. We will assume that the irreversible dynamics of the devices is sufficiently fast that at each time step they rapidly relax to new steady states. The occupation probabilities then adiabatically change from initial values to final values, $p_{x,i}\rightarrow p_{x,f}$. In the case of a sensor the bias forces are external to the CA, while for the case of an actuator the bias forces are internal to the CA.

According to the external description, the average energy and average entropy of a sensor and actuator will change over this time. The change in average energy is $\bar{E}=E(p_{1,f}-p_{1,i})$. For example, if $\epsilon_f=-\epsilon_0$ the change in average energy is 
\begin{equation}
\bar{E}= -E\tanh{\epsilon_0/2}~.
\end{equation}
In the case of a sensor $\epsilon_0 <0$ and the average energy increases, while in the case of an actuator $\epsilon_0 >0$ and the average energy decreases. In this example the average entropy does not change as the probabilities for each state are exchanged. Thus the change in the Helmholtz free energy is equal to the change in the average energy. This implies that work is done on the sensor while the actuator does work on the world.  

According to the internal description of the system, the internal state is a stochastic variable. The sensor spends most of its time in the low energy state and the actuator in a high energy state.  As the bias forces act this will change and the most likely state will switch. The time for system to switch is random variable: some change their state early in the control pulse and some change their state late in the control pulse. Some may not change at all. This implies that the work done by/on the external world, $w$, is in fact a random variable in the internal description.  

This scenario is typical of problems addressed in the field of stochastic thermodynamics~\cite{Seifert2012}. As bias forces change, and control pulses act over some time, the switching probabilities change.  Some of the key results in stochastic thermodynamics  relate the probability distribution for the work done to the changes between the initial and final stationary occupation probabilities. The surprising fact is that these relations can be independent of how the bias forces vary in time. Some examples for finite-state Markov models are given in \cite{Rao}.

As an example suppose that at the start of a control pulse, sensors and actuators are in the state of zero energy for a sensor and $E>0$ for an actuator.  This is most likely to be the case.   For an actuator, the probability for the actuator to change its state and  be found in the low energy state at the end of the control pulse is $p_{0,f}$, and the change in the internal energy is $-E$.  The probability that the device remains in its high energy state is the error probability  $p_e=p_{1,f}$. In that case the change in the internal energy is $0$. So  there are two possible values for the change in the energy of the system, $0$ and $-E$, with probability distributions, $Pr(0)=p_e$ and $Pr(-E)=1-p_e$. The work $w$ done by the system on the outside world is one of these two values, $0,-E$, fluctuating between the two values from trial to trial. Similar statements can be made for a sensor. The error probability is now $p_e=p_{0,1}$ and the the work done on the system is a random variable $w\in {0,E}$.

The Jarzynski equality \cite{Jarzynski} is a relation between the ensemble averaged values of $w$ over many trials and the change in free energy corresponding to the two distributions $p_x,p_x'$ for systems in contact with a heat bath. The equality thus relates the internal (stochastic) description to the thermodynamic external description. It is given by
\begin{equation}
    {\cal E}[e^{-\beta w}]=e^{-\beta \Delta F}~,
\end{equation}
where $\beta^{-1} =k_BT$ and $\Delta F=\Delta E-\beta \Delta S$, with $E$ the average energy and $S$ the average entropy (in natural units) for each of the distributions $p_x,p_x'$. In our presentation there is no requirement for the sensors and actuators to be thermal systems: they are maintained in arbitrary non-equilibrium steady states. Nonetheless a Jaryznski type equality holds (see~\cite{Seifert2012}). 

Sensors and actuators are not sufficient to define a causal agent.  There needs to be an interaction between the states of sensors and actuators and an internal learning machine. This is a physical irreversible process coupling the fluctuating energy states of the sensor/actuator, on the basis of time spent in the high/low energy state, to physical states of the learning machine. We will describe this in more detail in the next section. 

\section{Learning in a causal agent}
\label{learning-machine}

In our model the only data a CA has access to is the content of its sensor and actuator records. In order to learn causal relations we grant the machine some additional systems that can implement learning based on this data. In this section, we describe a model for this process.

Before we do so, however, let us first consider a simple, but thermodynamically expensive, way in which a CA could learn employing a device that uses `memory plus look-up'~\citep{RusNor2010}. In this model, the CA simply keeps its entire record of sensor and actuator data. In order to produce a particular sensation the CA scans the data until it finds the appropriate subset of actions to produce a sensation with high probability. This requires storing an immense amount of information and scanning it quickly. The average number of bits to store that data is given by the Shannon information of the record. If the functional relation to be learned is simple enough (a few bit Boolean function for example) this could be an effective strategy. In general however the function may require many real valued inputs, $f(x_1,x_2,\ldots x_n)$. In that case a very large number of trials will need to be stored: the number grows exponentially with $n$. This is known as the curse of dimensionality in machine learning. The point of learning is to compress this into a much smaller set of functional relations $f_w(x)$ where the long-time values of the weights label the functions that are learned. We will now describe how this can be done in an autonomous machine. 

\begin{figure}[t]
    \centering
    \includegraphics[scale=0.7]{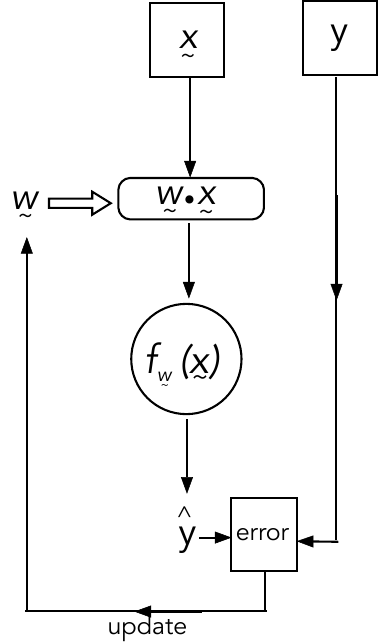}
    \caption{A very schematic representation of a machine learning algorithm. An element of the training data $\vec{x}$  is first summed over a set of weight parameters using a scalar product.  The result is then fed to a binary valued non-linear function to give the trial result $\hat{y}$. This is compared to the correct label $y$ and the values of the weight parameters are adjusted accordingly by an  update algorithm. The algorithm then repeats on the next element of training data.}
    \label{general-learn}
\end{figure}

We need to make a distinction between machine learning algorithms and machines that learn. The latter are computer programs that change the parameterisation of a set of functions of an extremely large number of variables, according to trial and error training. While the function itself is unknown, a sampled set of known values of the function $y_k= f(\vec{x}_k)$ is known. The inputs, $\vec{x}_k\in {\mathbb R}^n$, are the training data and $y_k$ is called the true label, for example, in binary classification $y_k\in \{0,1\}$. In the first trial, the algorithm computes a different function, parameterised by a set of real numbers $\vec{w}\in {\mathbb R}^M$,  $\hat{y}_k=f_{\vec{w}_k}(\vec{x}_k)$ on the training data. The value $\hat{y}_k$ is compared to the true value $y_k$ and if they are not the same the parameters $\vec{w}$ are changed according to some fixed algorithm (for example back propagation)  that ensures that the values converge probabilistically to the true labels.  In neural networks the functions are generated by nesting a large number of elementary functions, for example, sigmoidal functions. These are called activation functions. In Fig.(\ref{general-learn}) we give an example of the steps in the algorithm.

A learning machine is not an algorithm but an open irreversible dynamical system. We will assume for now it is described by a discrete dynamical map. The input is a set of signals $\vec{x}$ and $y$, which we call the training-signal and the label-signal respectively. The signals are physical quantities such as a voltage or a current. In each time step, the set of input training-signals, $\vec{x}$, is fed into a non-linear dissipative device that is biased using a set of physical parameters, the weights $\vec{w}$, for instance the bias voltages in an analogue circuit. The output-signal, here taken to be a single binary-valued physical quantity $\hat{y}$, is then compared to the label-signal and an error signal is used to feedback and control the weight settings according to a specific control scheme, which we will call `cooling', before the process repeats in the next time step.

Initially, the weights are randomly distributed, but as the machine evolves in time the distribution of weights converges to a narrow distribution on specific values that represent the unknown function $f_{\vec{w}}(\vec{x})$ inherent in the correlation between training-signal and label-signal.

Thus, we wish to make clear here that the essential difference between a machine learning (ML) algorithm and a learning machine (LM) is that a ML algorithm must be programmed into a suitable computational device, whereas a LM leverages the dynamics of its systems to learn -- nature does all the work. In particular: (i) the ML algorithm processes numbers, while the LM processes physical input signals; (ii) the ML algorithm acts on numbers through a conventional digital computational process running on a universal computer, while the LM is a physical machine with feedback control of irreversible dynamics; and (iii) the role of data and labels in a ML algorithm are played by the values of physical signals input to the LM, and model parameters such as weights by physical parameters that control the operation of the LM's internal dynamics. 

Our concept of a learning machine has more in common with analogue computers, such as the differential analyser from the middle of last century, than it does with the standard von Neumann paradigm for numerical computation. However unlike a differential analyser, friction, and the noise that accompanies it, are essential for the operation of a learning machine. Considering learning machines along these lines as a model of learning offers us a richer, and more biologically relevant, route to the design of machines that learn. It will also enable us to ground causal claims as relations between the physical states of learning machines. We will now describe how a machine can learn through a natural process to efficiently exploit thermodynamic resources. 

\section{Thermodynamic constraints on learning machines}
\label{therodynamics-learning}

\begin{figure}[t]
    \centering
    \includegraphics[scale=0.25]{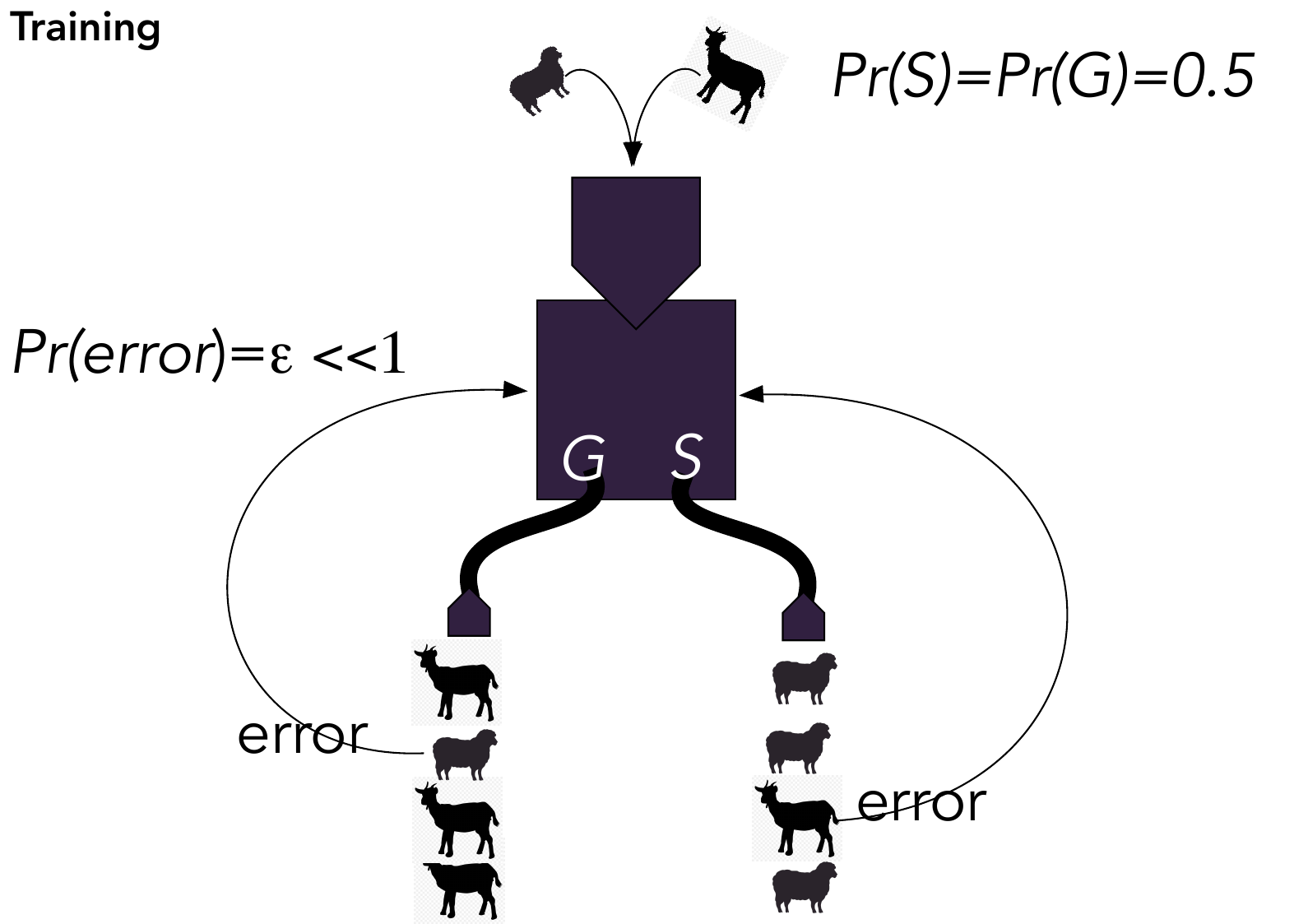}
    \caption{A learning machine is trained by feedback to classify images of goats and sheep. A correctly labelled prediction means the input image, chosen at random, comes out in the correct output.  At the end of a successful training run the probability of error, that is the probability that an image comes out in the wrong channel, is very small $Pr(error) =\epsilon <<1$. This is achieved by actively feeding back to the machine parameters every time a mistake is made.}
    \label{sheep-goats}
\end{figure}

Our goal in this section is to explicitly establish the link between learning in a machine and thermodynamics. To motivate the discussion we begin with a general argument that suggests why learning machines necessarily dissipate heat and generate entropy.

Suppose we desire a learning machine that will distinguish images of sheep from images of goats (Fig.(\ref{sheep-goats})). In the training phase a well-labelled image of a sheep or a goat is selected with equal probability and input to the machine. The output has two channels corresponding to the true labels of the input images. At the beginning, the image comes out in either channel with equal probability: roughly 50\% of the outputs match the true label and come out in the correct channel but about the same percentage do not. These are errors.

Whenever an error occurs, feedback conditionally changes the internal biases in the learning machine to try to decrease the error probability $Pr(error)=\epsilon<1$. Eventually the machine gets the true labels right and images almost always come out in the right channel. But it can never be perfect. To see this let us look at the entropy of the records at input and output 

The entropy at the input is clearly $\ln 2$ in natural units. Initially the output has the same entropy but, at the end of training, the entropy of the output records is reduced to the order of $\epsilon << \ln 2$.  This must be paid for by an overall increase in the entropy of the machine's environment through heat dissipation that arises every time work is done in the feedback steps. How does this arise in a physical learning machine?     

\begin{figure}[t]
    \centering
    \includegraphics[scale=0.5]{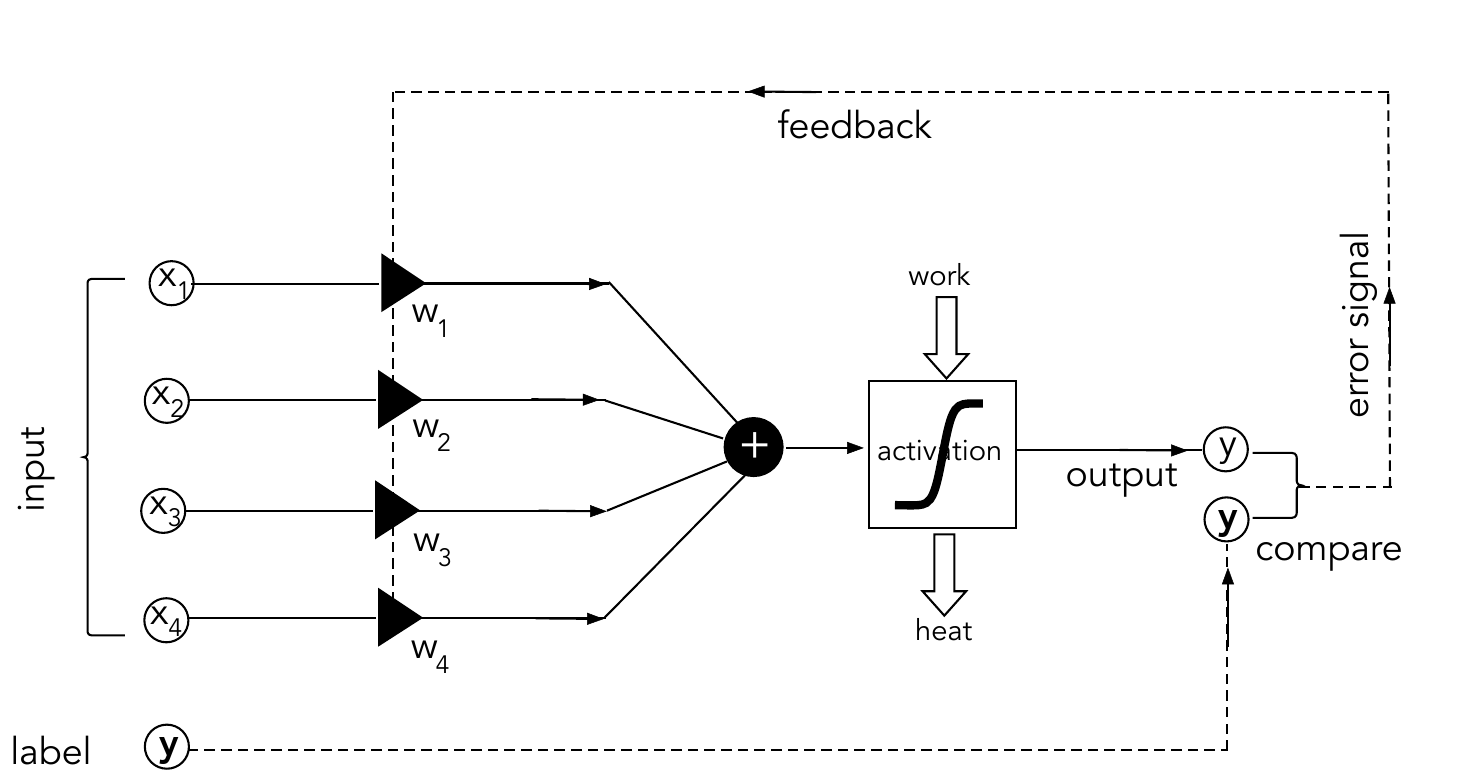}
    \caption{Schematic representation of an elementary learning machine, the perceptron, in training mode. A set of training data $x_n$ (real numbers) together with the correct label for that data ${\bf y}$ (binary number) are the input. The data $x_n$ are first multiplied by a weight factor $w_n$ (real numbers) before being summed to give $w.x= x_1w_1+x_2w_2+\ldots w_nx_n$ and passed to a non-linear device, the `activation switch', that produces an output signal $y=f(w.x)$ where $f$ is a binary valued non-linear function. This binary number is compared to the true signal-label ${\bf y}$. If they are the same, nothing is done but if they are different an error signal is sent to change the set of weights and repeat. Unavoidable physical noise means that the output is always subject to error. The cycle continues until this error probability is as small as possible. Outside of training mode the dashed control lines are removed. The device implementing the activation function is a work-driven and dissipative non-linear system. It must relax quickly to a steady state output (a fixed point or a limit cycle) when the input is changed.}
    \label{threshold}
\end{figure}

At the most elementary level neural network algorithms make use of a threshold or {\em activation function}. In learning machines the activation function becomes an {\em activation switch}: a dissipative, stochastic system with a non-linear relationship between input signals and output signals (Fig.(\ref{threshold})). The operation of the switch depends on setting bias parameters, similar to sensors and actuators. The biases represent physical states of the switch. The switch's bias parameters, which we refer to as weights, are changed by feedback in the process of learning. 

\begin{figure}
    \centering
    \includegraphics[scale=0.5]{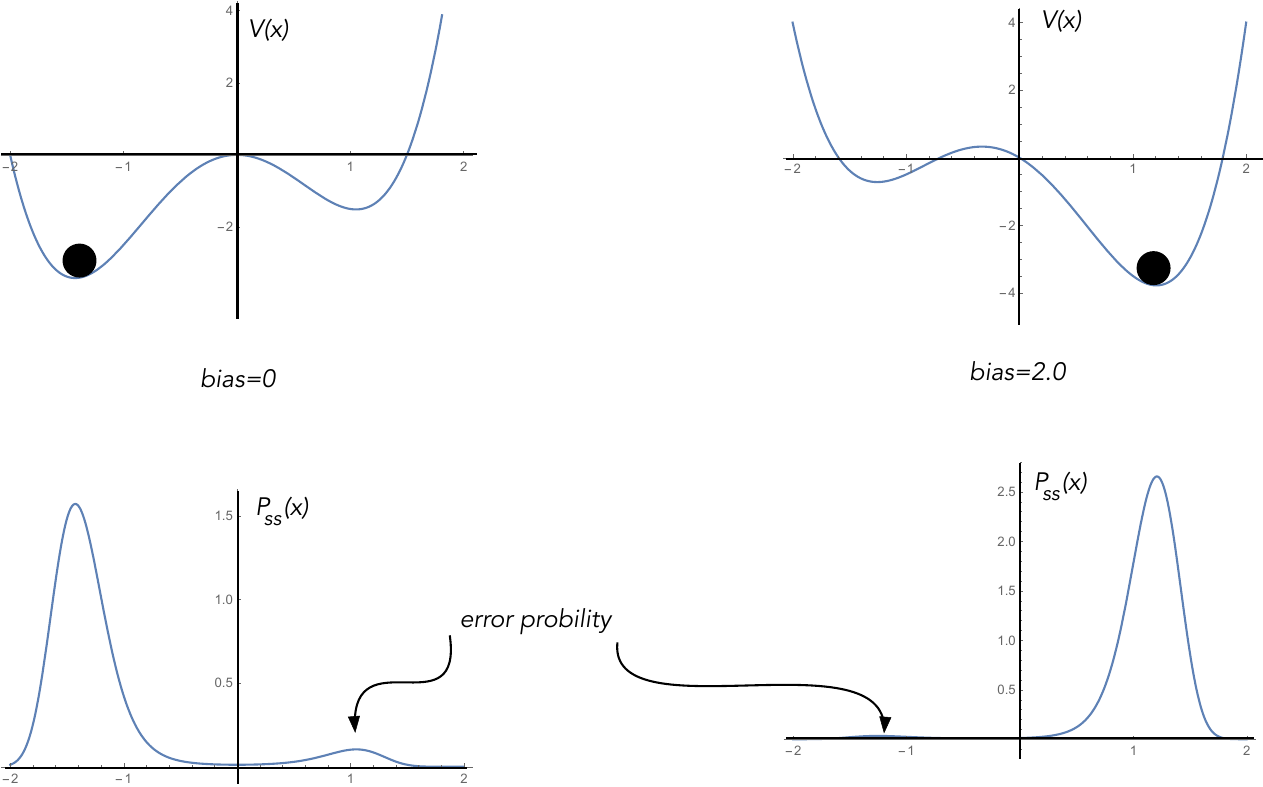}
    \caption{An example of an activation switch based on a particle moving with high friction in a double-well potential. The output variable $y$ is simply the sign of the displacement $x(t)$ of the particle. (Top) Application of a constant bias force changes the shape of the potential so that the lowest energy state changes from negative to positive displacement. This requires work to be done on the particle, work that is dissipated as heat. (Bottom) The steady state probability distribution when noise is included obtained by solving the Smoluchowski equation (see \citep[Ch.6]{Gardiner}). Note that there is a small probability of finding the system in the wrong position.}
    \label{double-well}
\end{figure}

An example of an activation switch is shown in Fig.(\ref{double-well}). A particle is moving with large friction so that its momentum can be adiabatically eliminated from the dynamics. Under these conditions the dynamics is effectively one dimensional and described by the Smoluchowski Fokker-Planck equation (see \cite{CP-LM} for details).

In a deep learning algorithm, weights parameterise functions and are changed according to the back propagation algorithm. In the case of a physical neural network, this becomes a physical feedback of signals from the output to the bias conditions. When we change the bias condition of a switch, we change the probability for the output to switch from one state to another. Changing the bias condition does work on the switch and that work is dissipated as heat. As for actuators and sensors, the work done and heat dissipated in each trial is a random variable. Averages over these random variables is constrained by the fluctuation theorems of stochastic thermodynamics~\cite{Seifert2012}.  
 
\begin{figure}[t]
    \centering
    \includegraphics[scale=0.5]{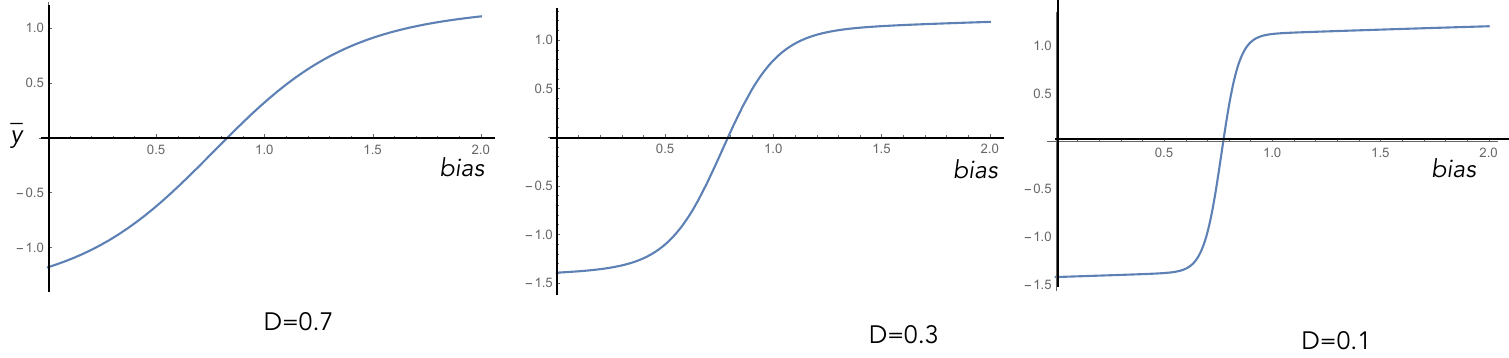}
    \caption{External description: The steady-state ensemble average of the mean displacement of the particle in the double well versus the bias with varying noise levels (i.e. temperature). On the left the noise is high and the activation like nature of the relationship between output and input is unclear. As the noise is decreased a more definite switch is seen as a function of the bias of the potential.}
    \label{DW-mean}
\end{figure}

\begin{figure}[t]
    \centering
    \includegraphics[scale=0.5]{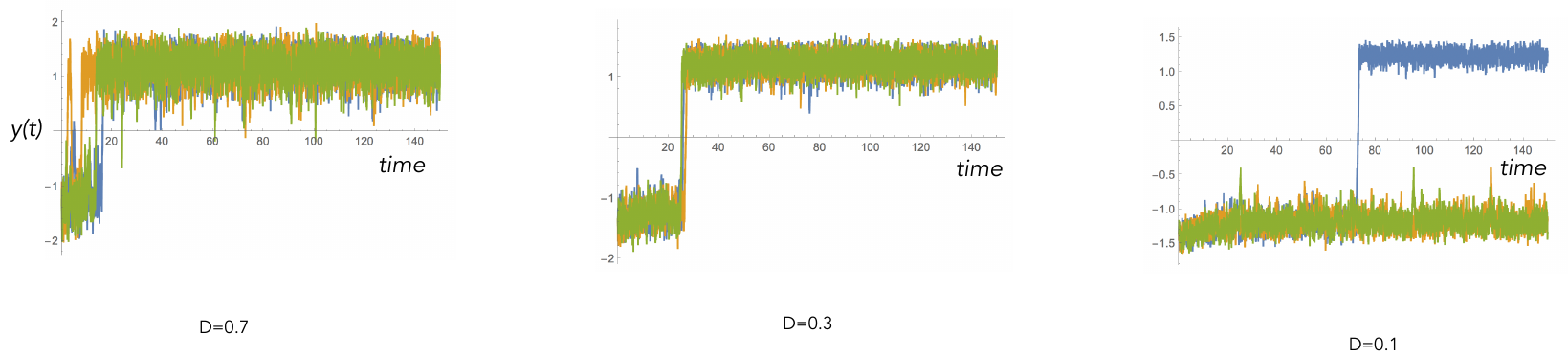}
    \caption{Internal description: Sample trajectories of the displacement of the particle in the double well activation switch as a function of time as the noise is decreased from left to right. In each case there are three samples $y(t)$. The bias of the potential was increased from $0$ to a constant maximum value. When the noise is large (left) the output switches quickly but the stochastic fluctuations are large. When the noise is small (right) in only one sample did the device switch at all implying an error on the two that did not switch. Furthermore, as the noise is decreased the switch is slower. In the limit of no noise there is no switching at all. Dissipation and noise are essential for efficient learning machines. (These plots were generated by solving the Ito stochastic differential equation for the Smoluchowski process in Mathematica.)}
    \label{DW-stochastic}
\end{figure}

Clearly such a system has a strongly dissipative non-linear dynamics and, by the fluctuation--dissipation theorem, must necessarily be intrinsically noisy. We must thus distinguish the ensemble average behaviour of the output -- as per the external description -- from the single trial output -- as per the internal description. This important distinction is explained in Figs.(\ref{double-well}), (\ref{DW-mean}), and (\ref{DW-stochastic}) with reference to a biased double well. In a single trial, the output of an activation switch is a random variable. This means that in some cases the output will not switch when it should, corresponding to an error (Fig.(\ref{DW-stochastic})). The objective of learning is to minimise this error by changing the biases/weights of the activation switch.

We contend that a further key feature that distinguishes a ML algorithm from a LM is the essential relation in a LM between the minimisation of error, and the minimisation of the thermodynamic cost of learning. Moreover, this relation connects an information theoretical concept -- error -- to a thermodynamic quantity -- heat dissipated. We set out a justification of this claim across the remainder of this section, where, following Goldt and Seifert~\cite{Seifert2017}, we base our discussion on the perceptron learning machine. We begin by describing a device that can take any number of input signals but produces a single binary signal at output. Our model is based on the two-state birth-death master equation system discussed in \S\ref{sensor-actuators}, and is thus intrinsically stochastic.

Consider a single system with one output signal labeled by $n\in\{-1,1\}$. This is a binary signal. The simple model in Fig.(\ref{bump}) can be described like this by setting $e_{n}=E(1+y/2)$. Suppose that there are $K$ binary input signals described by a data vector, $\vec{x}=(x_1,x_2,\ldots, x_k$) where $x_k\in \{-1,1\}$. We set the LM the objective to learn a binary-valued function of the $K$ input signals, $f(\vec{x})=n_T$, where $n_T$ is the true signal-label. The training data thus consists of the inputs $(\vec{x},n_T)$. As we discuss in more detail below, in the context of a CA, the inputs $\vec{x}$ to the learning machine are supplied as internal actuator records in the CA and the true signal-labels, $n_T$, are supplied by the world via the CA sensors.

There are $4^K$ such functions~\cite{Anthony} and there are many supervised machine learning algorithms that can achieve this, for example Valiant's PAC algorithm~\cite{Valiant}. However our interest is in designing a single machine that is able to learn any such function depending on the choice of $n_T$. To keep it simple we will discuss the case of $K=1$. There are four such functions; two constant functions, the identity function and the NOT function.  Let us consider the NOT function. The example discussed in Fig.(\ref{bump}) can easily be framed in terms of learning a binary function of a single binary variable. 

The two output states are described by the Markov process
\begin{equation}
  \frac{dp_1}{dt}= \mu(1-p_1)-\nu p_1~,  
\end{equation}
and $p_1+p_{-1}=1$. The transition rates, $\mu,\nu$, are non-linear functions of a weighted sum $\vec{w}.\vec{x}$ of the components of the training vector, $\vec{x}$, and the weight vectors $\vec{w}=(w_1,w_2,\ldots w_k)\in {\mathbb R}^K$. The training of the device is done by changing the transition rates between states by changing the weights $\vec{w}$ at each trial. If the two states represent, say, a coarse-graining of an underlying double well potential, the transition probabilities reflect thermal activation over a barrier depending on the bias forces applied~\cite{McW}. We will assume that the dynamical properties of the switch are such that whenever the weights change the transition probabilities, it rapidly relaxes to the new steady state probabilities given by 
\begin{equation}
\label{two-state-ss}
    \frac{p_1}{p_{-1}} =\frac{\mu}{\nu}~.
\end{equation}

We will choose $p_1(\vec{w})$ to model a particular kind of activation switch described by a function called a sigmoidal perceptron~\cite{RusNor2010}, 
\begin{equation}
\label{sigmoid}
 p_1(\vec{w})= \frac{1}{1+e^{-\beta E_0 \vec{w}.\vec{x}}}~,
\end{equation}
where $\beta$ is a constant fixed by experimental design. In a thermally activated device it would be $\beta^{-1}= k_BT$. In a quantum tunnelling device it is some function of tunnelling rates. Note that $\vec{w}.\vec{x}\in {\mathbb R}$ and can be positive or negative. 

At this point we pause to make a note of an important difference between a physical sigmoidal peceptron and a sigmoidal function used in machine learning algorithms. In the latter, the output is a real number while in the physical device the output is always a binary number, $\pm 1$. It is the probability distribution over these vales that is sigmoidal. However we can make a direct connection by evaluating the average output $\bar{n}(\vec{w})=p_1(\vec{w})-p_{-1}(\vec{w})$. Normalisation then implies that $\bar{n}(\vec{w})=2p_1(\vec{w})-1$, or $p_1(\vec{w})=(1+\bar{n}(\vec{w}))/2$. Thus we can sample $p_1(\vec{w})$ by running many trials with the same value of $\vec{w}.\vec{x}$ and recording the average output $\bar{n}(\vec{w})$. This set of trials is called an epoch. It is a time average over many identical trials. Another way to think of the averaging process is to imagine that we replace the single physical perceptron with a large number of identical perceptrons and only look at the average of $n$ over all of them. Whichever way we look at it, we will assume that the feedback control is based on computing the average output signal over many identical trials.  

Before we can proceed with a description of the feedback process that will be used to change the weights to  implement optimal learning we will consider the thermodynamics of the perceptron activation switch. This will enable us to define a thermodynamic cost function that controls the rate of learning.

When the weights change from one epoch to the next, what is the average work done on/by the device and what is the average heat dissipated? Let $\vec{w}_j, \vec{x}_j$ denote the weights and input signals at the $j-$epoch.  We will assume that the energies of each state of the two-state device are $e_{n}=E_0y$. This is equivalent to the simple model in Fig.(\ref{bump}) with a shift in the base-line energy. The change in the average internal energy between two successive epochs is~\cite{CP-LM}
\begin{equation}
\label{change-av-energy}
    \Delta \bar{E}= 2E_0\Delta \bar{p}_1(\vec{w})~,
\end{equation}
where $\Delta \bar{p}_1(\vec{w})=\bar{p}_1(\vec{w}+\Delta\vec{w})-p_1(\vec{w})$ and $\Delta\vec{w}$ is the change in the weight vector between two successive epochs. We will now impose the thermodynamic constraint that the objective of feedback is to change the weights in such a way as to minimise the average energy change per epoch. 

As it stands this does not easily compare with the usual way of implementing learning in which the focus is on minimising a cost function; for example, the average error between the output $\bar{n}(\vec{w})$ and the true label $n_T$. A simple way to relate the two approaches is presented in~\cite{CP-LM}.

We define the error per trial as
\begin{equation}
\label{random-error2}
\epsilon =\frac{1}{4}(n_T-n)^2=\frac{1}{2}(1-n_T n)~,
\end{equation}
where $n=\pm 1$ is the value of the random variable at the output of that trial. Averaging this over an epoch gives
\begin{equation}
\label{av-error-prob}
\bar{\epsilon}(\vec{w}) =\frac{1}{2}(1-n_T \bar{n}(\vec{w}))~.
\end{equation}
The change in the average error probability due to a variation in the weights is given by
\begin{equation}
\Delta \bar{\epsilon} =\Delta \vec{w}\cdot \vec{\nabla}_w\ \bar{\epsilon}~.
\end{equation}
We need this to decrease as much as possible per trial so we set
\begin{equation}
\Delta \vec{w}= -\eta \vec{\nabla}_{\vec{w}}\ \bar{\epsilon}~,
\end{equation} 
where $\eta$ is a positive scaling constant. Thus
\begin{equation}
\Delta \bar{\epsilon}=-\eta|\vec{\nabla}_{\vec{w}}\ \bar{\epsilon}|^2~.
\end{equation}
Hence, the feedback rule is  
\begin{equation}
\Delta\vec{ w}=\eta n_T\vec{\nabla}_wp_1(\vec{w})~.
\label{feedback-rule}
\end{equation}

With the choice of Eq.(\ref{sigmoid}), we find that 
\begin{equation}
\vec{\nabla}_w\  p_1= \beta p_1(1-p_1)\vec{\xi}
\end{equation}
and 
\begin{equation}
\label{delta-weight}
\Delta \vec{w}=\eta  \frac{n_T\beta(1-\bar{n}^2)}{4}\vec{\xi}~.
\end{equation}
The change in the weights depends on both the  training data and the corresponding true labels. Note that this goes to zero as $\bar{n}\rightarrow \pm 1$, corresponding to learning the required function. With this choice we find 
\begin{equation}
\Delta \bar{\epsilon}=-\eta\beta^2(1-\bar{n}^2)^2/16~.
\end{equation} 
This is always negative, and has a maximum when $\bar{n}=0, (p_1=1/2)$. At the start of training this is typically the case and the change in the average error is large. As training proceeds it decreases.
Using Eq. (\ref{change-av-energy}), we find
\begin{equation}
\label{energy-error}
     \Delta \bar{E}=-2\eta n_TE_0 \Delta \bar{\epsilon}~. 
\end{equation}
We see that the average change in energy per trial will be a minimum when the average change in error per trail is a minimum. Thus the cost function is equivalent to minimising a thermodynamic cost. This relates an information theoretical quantity, average error in learning, to a physical thermodynamic quantity, heat dissipation. This is similar to Landauer's principle~\cite{landauer} that relates the energy cost to an information theoretical quantity, decrease in entropy by erasure of information.  A different thermodynamic constraint on learning is given by Goldt and Seifert~\cite{Seifert2017}. They show that the mutual information between the true labels and the learned labels is constrained by heat dissipated and entropy production.     

The discrete dynamical process induced on the space of weights by feedback is stochastic because in each epoch the output probability distribution needs to be sampled (Eq.(\ref{delta-weight})). However this is a highly non-linear stochastic process. When learning is complete the weights reach a new stationary distribution on weight space that has much less entropy than the distribution used at the start of learning.  

If we initialise the weights by a random vector, then the average output over an epoch, $\bar{n}(\vec{w})$ is close to zero and the average error is large. The steady state distribution of the preceptron is uniform and its entropy is a maximum. The learning proceeds by feeding back onto the $\vec{w}$ until $\bar{n}(\vec{w})^2\approx 1$. We can also relate the change in average entropy of the perceptron per epoch to the change in average error per epoch~\cite{CP-LM}. The result is that the decrease in average entropy per step is also proportional to the change in average error per step.

Stepping back, it is clear that a learning machine is a dissipative driven non-equilibrium system with a highly non-linear dynamics and many variables.   Betti and Gori~\cite{BettiGori} have made a similar claim. Like all such systems learning machines have non-equilibrium steady states and corresponding basins of attraction in weight space~\cite{Seifert2012}. Initial high entropy distributions over weight space are `cooled' into the basins of attraction that characterise the learned function. The learning is driven by an imperative to optimise the use of thermodynamic resources. Where does that imperative originate? It might be the case that in an open far-from-equilibrium system of sufficient complexity the evolution of learning machines is a direct consequence of thermodynamics and evolution~\cite{Rovelli2016,Rovelli2019,England}.

The function that the machine learns is labelled by the weights that it converges to in training. These are not deterministic but, on the external description, the weights after training are sharply peaked around particular values that label the function learned, $f_{\vec{w}}$. These labels are physical variables that define the bias parameters of the learning machine. This is how learning machines come to represent functions in terms of physical variables.

\section{Learning and causal relations}
\label{learning-causes}

\begin{figure}[t]
    \centering
    \includegraphics[scale=0.6]{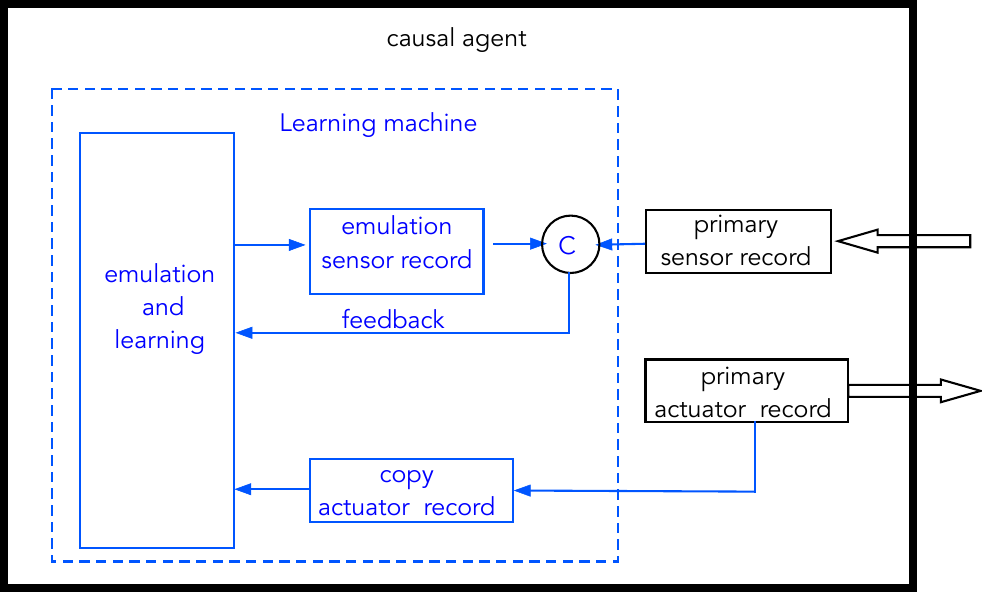}
    \caption{A schematic of a learning machine based on a physical emulator with feedback.  A single round of learning proceeds as follows. The primary actuator record registers what action is taken on the external world while the primary sensor record indicates what sensation is received from the external world. The primary actuator record is copied and sent to an emulation engine to produce an emulated sensor record. This is compared to the primary sensor record by a comparator (C) and the result fed back to the emulation engine which then updates. A new action is taken and the process repeats until some goal is met for the comparator output. The feedback process and update may be a discrete time or continuous time stochastic process. This model for learning is similar to the concept of predictive processing recently developed in the philosophy of neuroscience (see~\cite{Kirchhoff2018} and references therein.)}
    \label{learning-emulator}
\end{figure}

We now return to the discussion of a causal agent that incorporates sensors, actuators, and learning machines. The model, which we refer to here as the emulator model (EM), is summarised in Fig.(\ref{learning-emulator}). This model is inspired by models now common in neuroscience that treat the brain as primarily a prediction machine~\cite{Clark,Seth,Buzsaki}. The core idea originated in the corollary discharge model of Sperry~\cite{Sperry1950}, but it is a general model for a learning machine. 

The goal of the learning machine is to take actuator records and predict sensor records. The predictions are then compared to actual sensor records and feedback is used to modify the settings of the learning machine dynamics so as to minimise the probability that predictions do not match actual sensor records. As we have already seem, this abstract informational goal can be made equivalent to a physical goal: minimise thermodynamic cost as measured by power dissipated by the CA as it interacts with its environment. 

There are no doubt many ways to engineer a learning machine along these general lines. We will use the physical neural network (PNN) model described in the previous section. In terms of that model, the actual sensor records coming from the external world play the role of true labels in supervised learning. We can think of nature as a function oracle to which actuators pass arguments, in the PAC language of Valiant~\cite{Valiant}. The input to the PNN emulator are the actuator records that record sequences of interventions taken by the machine on the external world. The outputs of the PNN are predicted sensor records and these are compared to the true labels provided by the actual sensors acted on by the world. The previous discussion shows that minimising error predictions is equivalent to minimising the thermodynamic cost of the CA interacting with the world. 

In this model, we take the correlations between sensor and actuator records learned by the CA to be causal relations. Thus, a causal relation is a learned function, $f_w(a)$, mapping actions to predicted sensations and parameterised by the physical quantities that represent the weights, $w$, inside the learning machine. As an example, let us return to the model in Fig.(\ref{bump}). It is clear that the function to be learned is a one bit Boolean function. Two of these are constant functions, that always give either 1 or 0. These give the same result regardless of the input. In terms of the physical model this would correspond to a bump that is high enough to reflect every particle or low enough to transmit every particle. The other two are the identity function and its complement, the NOT function. In both of these the output depends on the input. Up to a relabelling of the states of the source, these are the same physical process. In particular, they are the only ones that can be the basis of an effective strategy (to use Cartwright's term) as far as the CA is concerned. 

\begin{figure}[t]
    \centering
    \includegraphics[scale=0.5]{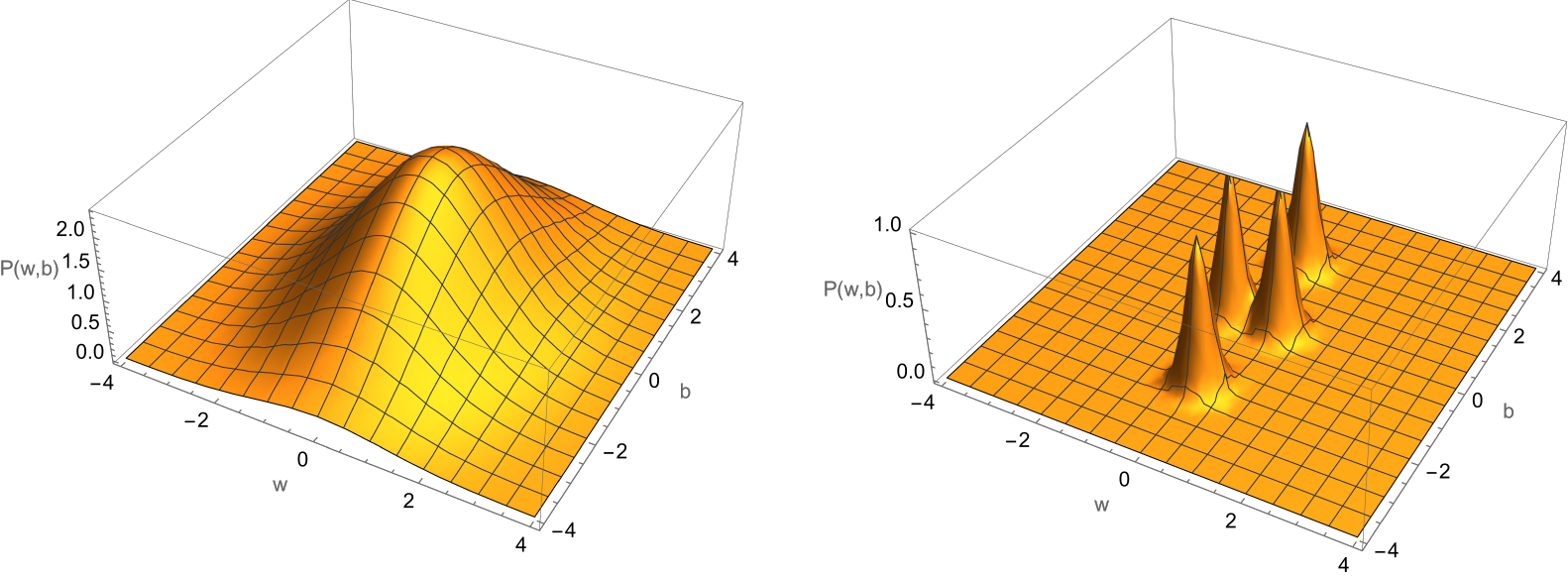}
    \caption{Learning a one bit gate with a single physical perceptron. Left: The initial distribution of the weight and bias variables. Right: The four possible distributions after training corresponding to the four different labels for the four possible one bit functions. Which one the machine learns depends on the training data labels.  For example, if the training data has a fixed label $n_T=1$ for all possible values of the input, $x$, the perceptron will stochastically evolve to $w=0,b>0$ with high probability.}
    \label{one-bit-weights}
\end{figure}

A one bit Boolean function can be learned by a single perceptron~\cite{CP-LM}. The probability function is given by a single weight $w$ and single bias $b$,
\begin{equation}
 p_1(w,b)= \frac{1}{1+e^{-\beta E_0 (w.x+b)}}~.
\end{equation}
The not function corresponds to $w<0,b=0$ and the identity function to $w>0,b =0$. The two constant functions correspond to $w=0, b \neq 0$. In Fig.(\ref{one-bit-weights}) we schematically indicate the four possible distributions for the weights ($w$) and bias ($b$) after learning has reached minimum error.  The initial distribution of the $w,b$ is very broad. After learning they are localised in the space as four distinct distributions, corresponding to the functions learned. 

It is important to keep in mind that the physical machine `knows' no more about the causal relations than the settings of its weights and bias. These label the function learned in a probabilistic sense: there is not a unique label only a region of likely labels for a given function. Can one build a learning machine that learns which function the machine has learned?

To answer this question we need to introduce the idea of a nested hierarchy of learning machines, wherein each level learns some feature of the functions learned at the lower level by learning some feature of the weights learned at the lower level, for example which of the four regions in Fig.(\ref{one-bit-weights}) have been reached. 

Suppose we have a learning machine with two actuators and one sensor. This could learn an arbitrary two bit function with a single binary output. There are 16 such functions. These functions could be classified by a property, and that property can be learned. For example, suppose the classification is simply the answer to the question, `is the learned function balanced or constant?' A balanced function is one that takes the same value on half of the input domain and zero on the other half. An example of this is XOR. 

The 16 possible functions can be labelled by a binary string of length 4. There are 256 binary functions with four inputs so the higher learning machine will need more perceptrons or more layers or both. It is know that a binary valued function of $n$ binary variables can be learned with a single hidden layer of $2^n$ perceptrons. There will be a large number of weights and biases to be learned in a back propagation. 

Thermodynamic constraints will continue to apply when learning machines are nested in this way.  At each level, optimal learning will correspond to optimal thermodynamic efficiency given a particular cost function. What sets the cost function? To answer this question we need to ask; why learn anything at all?

One answer that is relevant for an interventionist characterisation of causation~\cite{Pearl2000, Woodward2003} is the following: a machine learns causal correlations between sensations and actions so as to act effectively in the world. An effective learning strategy as one that enables effective control of some system external to the CA. This insight shifts the question somewhat: what is an effective control strategy? This question is addressed in the field of optimal stochastic control~\cite{Stengel}, which explicitly addresses the question of what is an optimal control {\em policy}, that is to say, how best to act.

The simplest example of an optimal policy would seek to minimise the energy used to change the state of a physical system~\cite{Stengel}. It seems likely that, in an evolutionary setting, such a cost function will be selected spontaneously. An old  result of cybernetics is the {\em good controller theorem}~\cite{good-regulator}, according to which the best stochastic controller is a replica of the system to be controlled. In our model of emulation learning, the learning machine does indeed seek to replicate the dynamics of the external world, in a simpler device, so far as it can see it through actions and sensations.  As we discussed in \S\ref{therodynamics-learning}, the free energy cost per learning step is minimised as the error change per step is minimised. This enables a physical grounding of the free energy principle of Friston~\cite{Friston}. Learning machines are thermodynamically optimal controllers of the external world and the cost function is ultimately thermodynamic in origin. It is not imposed from the outside. Effective strategies are grounded in thermodynamics.

\section{Discussion and Conclusion}
\label{conclusion}

Let us finally return to the question of causal asymmetry in the context of our approach. A response to this question crucially depends upon whether causation is an objective feature of the world, or whether it is grounded in the perspective of certain special kinds of physical systems, such as ourselves, that learn and act. One of our primary goals in this work has been to demonstrate how causal relations can be defined in terms of the physical states of a special kind of physical machine, a casual agent. The sole purpose of such physical states is to enable prediction of the sensations that follow an action, within some error bound. Thus we take these physical states, which are characterised by the internal weights and biases of the learning machine of the CA, to encode a custom-built causal model of the CA's environment (where the internal weights and biases are simply the modelling parameters of the causal model).

On this view the causal relations {\em just are} the internal settings of the physical states of the learning machine, which are no less physical than those in the world outside. As such, we have argued that causal relations are nothing more than learned relations between sensor and actuator records inside the learning machine of the CA. What is more, we have demonstrated the close connection between the operation of a physical learning machine, and so also the causal agent, and the laws of thermodynamics and energy dissipation. We argue that efficient learning, in the sense of minimising error, is equivalent to thermodynamic efficiency, in the sense of minimising power dissipation. Along with the thermodynamics of sensors and actuators, this renders any such causal agent inherently `directed', and we take this fact to underpin the asymmetry of causation.

We treat the consequences of our approach in greater philosophical depth in \cite{ESM}. Briefly, however, let us outline the way in which we take the above considerations to exemplify a perspectival approach to the interventionist account of causation~\cite{Price2007,Ismael2015}.

Our account is explicitly interventionist due to the necessity of actuators in a CA. These are physical subsystems in a CA that do work on the external world. Sensors alone are not enough to build a CA. Certainly one could easily build a CA with an algorithm to find patterns in its sensor records (a Bayesian network, say), especially with time-series data from multiple sensor types. It is easy to see that correlations could be found between records from different types of sensors. But would any extant correlation in the records indicate a causal relation? Such a claim would be open to Hume's objection: patterns or `regularities' do not ground causal claims. It is only through intervening on the world that agents can confidently discriminate cause from effect.

Moreover, our model of a CA demonstrates how the asymmetry between exogenous and endogenous variables in the interventionist account of causation has its origin in the thermodynamic asymmetry of the CA itself. Since the causal model that the CA learns is underpinned by its own mechanisms of intervention, detection, and learning, then the thermodynamic directedness of its actuators, sensors, and learning machine dictate that its causal model must also be `directed' in the same way. That is, the CA must be modelling its environment with itself embedded at the heart of the model, such that effects in the world must be thermodynamically downstream from its actuators and thermodynamically upstream from its sensors, and any modelled causal relations must be directed from variables manipulable by the actuator (exogenous), and detectable by the sensor (endogenous). Thus, in a sense, this directedness constrains the agent to be able to act only towards the `thermodynamic future', and to gain knowledge only of the `thermodynamic past'.

We take this to be a clear demonstration of what Price~\cite{Price2007} calls a causal perspective: agents who are situated and embedded `in time' in this way are constrained with respect to the actions they can perform and the functional relations they can exploit. Moreover, it is clear that this perspective is inescapable by the CA -- it is a function of its own internal constitution. What is more, a CA is also constrained to model its environment exclusively in terms of the set of dynamical variables that are manipulable-and-detectable by the actuator and sensor system. As such, the causal model learned by the CA will only contain functional relations that are exploitable according to the physical constitution of its sensors and actuators. This is then a further sense in which the CA is situated and embedded in an inescapable environment, again as a function of the physical constitution of its own network of actuators, sensors, and learning machines. There is an obvious connection between this latter situatedness and Kirchoff's~\cite{Kirchhoff2018} notion of an agent's own `Markov blanket' as an environment of sorts with which its internal states must contend. Once again, this situatedness defines a causal perspective.

As a final speculative suggestion, we wish to note that CAs that inescapably share a functionally similar constitution of actuators, sensors, and learning machines will share a `perspective' with respect to the causal models they learn. Thus a shared physical constitution ensures the stability of a casual perspective across a class of CAs, and this then defines an equivalence class of agents that shares a perspective. It is in this way, then, that learned causal relations are shared by all CAs with the same hardware embedded in the same environment. The ramifications of this sort of multi-agent learning we think is ripe for further investigation.

Our approach here shows that we can define causal relations independently of human agency, giving a perspectival interventionist account that avoids the charge of anthropocentrism. Instead we define causal relations as learned relations between internal physical states of a special kind of open system, one with special physical subsystems -- actuators, sensors, and learning machines -- operating in an environment with access to a large amount of free energy. Even though such open systems need not necessarily be human, it is their ability to model the causal relations in their environment, and exploit such relations as effective strategies, that make them causal agents.
  
\section*{Acknowledgements.}
This work was supported by FQXi FFF Grant number FQXi-RFP-1814 and the Australian Research Council Centre of Excellence for Engineered Quantum Systems (Project number CE170100009).  We thank Ken Arthur, Jenann Ismael, Michael Kewming, and Huw Price for useful discussions.


\begin{thebibliography}{99}

\bibitem{Russell1913}Russell, B. (1913). {\em On the Notion of Cause}, Proceedings of the Aristotelian Society,
13: 1–26.

\bibitem[Pearle (2000)]{Pearl2000}Pearl, J. (2000). {\em Causality: Models, reasoning, and inference}, Cambridge: Cambridge University Press.

\bibitem[Woodward (2003)]{Woodward2003}Woodward, J (2003), {\em Making Things Happen: A Theory of Causal Explanation}, New York: Oxford University Press.

\bibitem{Ismael2015}Ismael, J., (2015) {\em How do causes depend on us? The many faces of perspectivalism}, Synthese, DOI 10.1007/s11229-015-0757-6

\bibitem[Price (2007)]{Price2007} Price, H. (2007)  {\em Causal perspectivalism}.  In H. Price and R. Corry (Eds.), Causation, physics, and the
constitution of reality: Russell’s republic revisited, (pp. 250–292) Oxford: Oxford University Press. 

\bibitem[Cartwright (1979)]{Cartwright1979} Cartwight, N., (1979) \emph{Causal Laws and Effective Strategies} Noûs, Vol. 13, No. 4, Special Issue on Counterfactuals and Laws (Nov., 1979), pp.419-437.

\bibitem[Ismael (2016)]{Ismael}Ismael, J. (2016) \emph{How Physics Makes Us Free}, Oxford University Press. 

\bibitem[Milburn and Basiri-Esfahani (2022)]{CP-LM}G. J. Milburn  and  Sahar Basiri-Esfahani, {\em The physics of learning machines}, Contemporary Physics, 63:1, 34-60.
  
\bibitem[Goldt and Seifert (2017)]{Seifert2017}Goldt, S and Seifert, U. (2017) {\em Stochastic Thermodynamics of Learning}, Phys. Rev. Letts.  118, 010601.
 
\bibitem[Russell and Norvig (2101)]{RusNor2010}Russell, S.J. and Norvig, P. (2010), {\em Artificial Intelligence: A modern approach.}, Prentice Hall New Jersey. 

\bibitem[Briegel, (2012)]{Briegel}Briegel, H.J., and De Las Cuevas, G., (2012) {\em Projective simulation for classical and quantum learning agents}, Sci. Rep. 2, 400, (2012)

\bibitem[Floridi (2011)]{Floridi2011}Floridi, L., (2011){\em The Philosophy of Information} Oxford University Press.

\bibitem[Friston and Stephan (2007)]{Friston}Friston, K., and Stephan, K. E. (2007). {\em Free-energy and the brain.} Synthese, 159(3), 417–458.\textbf{}

\bibitem[Bruineberg, et al. (2018)]{Bruineberg} Bruineberg, J.,  Kiverstein, J. and Rietveld, E.,{\em The anticipating brain is not a scientist: the free-energy principle from an ecological-enactive perspective}, Synthese (2018) 195:2417–2444.

\bibitem[Freeman (2008)]{Freeman}Freeman W. J. {\em Perception of time and causation through the kinesthesia of intentional action}, Integr. Psychol. Behav. Sci. 2008 Jun;42(2):137-43.

\bibitem[Carmichael (2008)]{Carmichael}Carmichael, H. J. (2008), {Statistical Methods in quantum Optics, VII}, Springer. 

\bibitem[Rao and Esposito (2018)]{Rao}Rao, R. and  Esposito, M. (2018), Detailed Fluctuation Theorems: A Unifying
Perspective, Entropy, 20, 635.

\bibitem[Jarzynski (1997)]{Jarzynski}Jarzynski, C. (1997) {\em Nonequilibrium equality for free energy
differences} Phys. Rev. Lett. 78 2690.
 
\bibitem[Seifert (2012)]{Seifert2012}Seifert, U. (2012) {\em Stochastic thermodynamics, fluctuation theorems and molecular machines}Rep. Prog. Phys. 75 126001

\bibitem[Anthony (2005)]{Anthony}Anthony M. (2005){\em Learning Boolean Functions} Centre for Discrete and Applicable Mathematics, LSE. CDAM-LSE-2005-24 (2005).
 
\bibitem[Valiant (2915)]{Valiant}Valiant, L. (2013), {\em Probably Approximately Correct: Nature's Algorithms for Learning and Prospering in a Complex World } Basic Books. 
 
\bibitem[McNamara and Wiesenfeld (1989)]{McW}McNamara, B. and Wiesenfeld, K., (1989) 
{\em Theory of stochastic resonance}, Phys. Rev. A,39,4854.

\bibitem[Landauer (1986)]{landauer}R. Landauer, (1986) {\em Computation and physics: Wheeler's meaning circuit? },Foundations of Physics , {\bf  16}, Issue 6, pp 551–564.

\bibitem{BettiGori}Betti, A., Gori, M., (2016), {\em The principle of least cognitive action} Theoretical Computer Science 633 (2016) 83–99.

\bibitem[Rovelli (2016)]{Rovelli2016}Rovelli, C. (2012) {\em Meaning = Information + Evolution}, arXiv:1611.02420v1.
 
\bibitem[Pollack et al. (2019)]{Rovelli2019}Jeffery,K., Pollack,  R., Rovelli, C., (2019) {\em On the statistical mechanics of life: Schrödinger revisited}, arxiv.org, 1908.08374
 
\bibitem[England (2015)]{England}England, J.L. (2015){\em Dissipative adaptation in driven self-assembly}, Nature Nanotechnology,  DOI: 10.1038/NNANO.2015.250
 
\bibitem[Kirchhoff (2018)]{Kirchhoff2018}Kirchhoff, M.D., (2018), {\em Predictive processing, perceiving and imagining: Is
to perceive to imagine, or something close to it? }, Philos.  Stud. 175:751–767
 
\bibitem[Clark (2013)]{Clark}Clark, A. (2013) Whatever next? Behavioural and Brain Sciences, 36, 181–253 \url{doi:10.1017/S0140525X12000477}

\bibitem[Seth (2022)]{Seth}Seth, A., (2022), {\em Being You: A New Science of Consciousness}, Faber \& Faber. 

\bibitem[Buzsaki (2019)]{Buzsaki}Buzsaki, G.,(2019), {\em The Brain from Inside Out},  Oxford University Press. 

\bibitem[Sperry (1950)]{Sperry1950}Sperry R. W.  (1950). {\em Neural basis of the spontaneous optokinetic response produced by visual inversion}. Journal of Comparative and Physiological Psychology. 43 (6): 482–9.

\bibitem[Llinas (2002) ]{Llinas}Llin\`{a}s, R. R. (2002) {\em I of the Vortex, from Neurons to Self.}. (MIT Press Boston)

\bibitem[Gardiner (1983)]{Gardiner}Gardiner, C.W. (1983), {\em Handbook of Stochastic Processes for Physics, Chemistry and the Natural Sciences}, Springer. 
 
\bibitem[Stengel (1993) ]{Stengel}Stengel, Robert  F., (1993), Optimal control and estimation,  Dover.
 
\bibitem[Conant and Ashby (1970) ]{good-regulator}Conant, R. C.  and Ashby, W. R., (1970), Every good regulator of a system must be a model of that system, Int. J. Systems Sci., 1, 89.

\bibitem[Evans et al. (2022) ]{ESM}Evans, P. W., Shrapnel, S., Milburn, G. J. (2022), Causal asymmetry from the perspective of a causal agent. \url{http://philsci-archive.pitt.edu/18844/}

\end{thebibliography}
\end{document}